TABLE OF CONTENTS

**Nonvolatile ferroelectric control of topological states in 2D heterostructures**

Hua Bai[1], Xinwei Wang2, Weikang Wu[3], Pimo He[1], Zhu'an Xu[1], Shengyuan A. Yang[3] and Yunhao Lu[1]*

[1] Zhejiang Province Key Laboratory of Quantum Technology and Device, Department of Physics, Zhejiang University, Hangzhou 310027, China

[2] State Key Laboratory of Silicon Materials and School of Materials Science and Engineering, Zhejiang University, Hangzhou 310027, P. R. China

[3] Research Laboratory for Quantum Materials, Singapore University of Technology and Design, Singapore 487372, Singapore

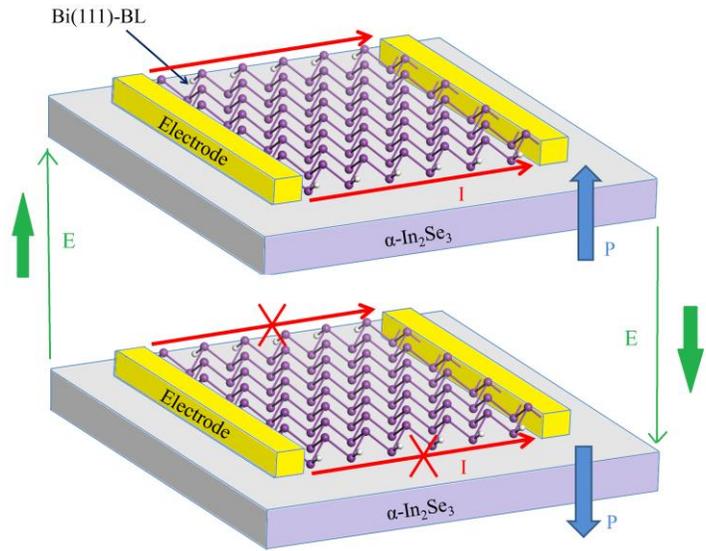

Predicted nonvolatile control of topological states in Bi(111)-BL/α-In$_2$Se$_3$ heterostructures.

# Nonvolatile ferroelectric control of topological states in 2D heterostructures

Hua Bai[1], Xinwei Wang[2], Weikang Wu[3], Pimo He[1], Zhu'an Xu[1], Shengyuan A. Yang[3] and Yunhao Lu[1] (✉)

[1] Zhejiang Province Key Laboratory of Quantum Technology and Device, Department of Physics, Zhejiang University, Hangzhou 310027, China

[2] State Key Laboratory of Silicon Materials and School of Materials Science and Engineering, Zhejiang University, Hangzhou 310027, P. R. China

[3] Research Laboratory for Quantum Materials, Singapore University of Technology and Design, Singapore 487372, Singapore

Address correspondence to Yunhao Lu: luyh@zju.edu.cn

## Abstract

Quantum spin Hall (QSH) insulator materials feature topologically protected edge states that can drastically reduce dissipation and are useful for the next-generation electronics. However, the nonvolatile control of topological edge state is still a challenge. In this paper, based on first-principles calculations, the switchable topological states are found in the van der Waals (vdW) heterostructures consisting of two dimensional (2D) Bi(111) bilayer (BL) and α-$In_2Se_3$ by reversing the electric polarization of the ferroelectric α-$In_2Se_3$. The topological switching results from the different charge transfer associated with the two opposite polarization states of α-$In_2Se_3$. This new topological switching mechanism has the unique advantages of being fully electrical as well as nonvolatile. Our finding provides an unprecedented approach to realize ferroelectric control of topological states in 2D materials, which will have great potential for applications in topological nanoscale electronics.

# 1 Introduction

2D materials have attracted immense attention in the past decade due to their fascinating properties and broad application prospects. It is easy to pattern 2D materials through chemical or mechanical techniques due to their atomic thickness. Meanwhile, the weak vdW interaction between adjacent layers offer unprecedented opportunities to form vdW heterostructures and explore quantum electronics at the nanoscale. Particularly, novel topological electronic states can be realized in 2D materials. The topological insulator (TI) in 2D system, also known as the QSH insulator, features an insulating bulk and conducting edge states that are topologically protected from backscattering by the time-reversal symmetry[1-5]. The mobile electrons at the topological edge states is spin–momentum locked, which will be extremely useful for applications in spintronics. Comparing with 3D TI, the edge states in 2D TI are more robust against back scattering because only counter-propagating and oppositely spin-polarized conducting states exist at the edges. To make useful device applications with 2D TIs, a prerequisite is to achieve controlled on/off switching of the topological edge channels. As the edge states is determined by the topology of the bulk electronic bands, the switching should involve topological phase transitions of the 2D system. Two approaches have been proposed in previous research: one is by strain[6-7] and the other is by gate electric field. The strain approach has been suggested to be viable for $Na_3Bi$ and $Cd_3As_2$ thin films and the device concept of piezo-topological transistor has been proposed[8]. However, the precise and reversible control of strain at nanoscale is still challenging in practice. The gate-field approach has been proposed for Dirac semimetal thin films[9], 1T'-MoS2 derivatives[10], and Bi films[11-12]. The recent experiment on $Na_3Bi$ thin films indeed demonstrated the possibility of topological switching[13]. Nevertheless, the required field strength is still quite large (> 0.1 V/Å), making it difficult for practical applications. Furthermore, an important feature of both approaches is that they are volatile, meaning that there must be constant energy input to sustain (either one of) the switched state. This will be a crucial disadvantage for nonvolatile device applications, such as nonvolatile memories.

Ferroelectric (FE) materials are a class of materials with spontaneous and switchable electric polarization. A natural idea is to achieve topological switching by combing a QSH insulator with a FE material. However, for 3D FE materials, interfacial strain due to lattice mismatch and dangling bonds between the QSH film and the FE substrate are typically unavoidable, which will dramatically change the electronic structure beyond the control of topological character. Recently, 2D FE materials have been predicted and some of them were realized in experiment, such as group-IV monochalcogenides monolayers[14], $Bi_2O_2Se$ family[15], group-V monolayers[16], few-layer Te[17], α-$In_2Se_3$ family[18], and 1T'-$MoS_2$ monolayer[19]. It is then possible to form vdW heterostructures with QSH materials and 2D FE materials. The vdW interaction between them can help to avoid the problems from lattice mismatch and dangling bonds. Therefore, such systems could offer new possibilities to achieved controlled topological switching.

In this paper, we demonstrate such a new approach. Based on first-principles density-function theory (DFT) calculations, we investigate the vdW heterostructures consisting of typical QSH insulator, Bi(111)-BL film and 2D FE materials，α-$In_2Se_3$. We show that the topological character of the heterostructure can be switched by switching the polarization direction of α-$In_2Se_3$. The topological switching is confirmed by detailed partial band structures analysis, the explicit calculation of the bulk $Z_2$ topological invariant, and the presence/absence of the topological edge states. We find that switching is connected with the different charge transfer between Bi(111)-BL and α-$In_2Se_3$ in the two different polarization states. Our work demonstrates a new approach of topological switching controlled by FE polarizations in 2D with the advantage of non-volatility, which will be useful for future topological electronics applications.

## 2 Computational methods

In this work, our calculations were performed using the Vienna ab initio simulation package (VASP)[20] with the projector-augmented wave (PAW) method[21]. The generalized gradient approximation (GGA) with Perdew, Burke, and Ernzerhof (PBE)[22] realization was used for the exchange-correlation functional. Hybrid functional (HSE06)[23] method was also employed to verify all band structure results. The vacuum layer more than 20 Å perpendicular to the sheets was employed to avoid the unphysical interactions between images caused by the periodic boundary condition. The energy cutoff was set above 400 eV and the force and energy convergence criteria were set to 0.01 eV/Å and $10^{-5}$ eV respectively. A $9 \times 9 \times 1$ Γ-centered k-point mesh was used for the Brillouin zone sampling for slabs, and $9 \times 1 \times 1$ for ribbons. The vdW corrections was included by the DFT-D3 method[24]. The $Z_2$ number and the edge states were calculated by the Wannier90 package[25] and the WannierTools code[26]. The irreducible representations were calculated by the Irvsp code[27]. The spin-orbital coupling (SOC) was considered in all calculations. The calculated equilibrium in-plane lattice constant of Bi(111)-BL and α-$In_2Se_3$ monolayer are 4.33 Å and 4.1 Å respectively, in agreement with the previous works[28-29]. Here, a = 4.1 Å was taken for the heterostructures, and the band topology of Bi(111)-BL does not change under such a small strain (see Fig. S1).

## 3 Results and discussion

Graphene was proposed as the first QSH insulator[30]. However, the SOC strength in graphene is too small to make the topological gap detectable. This has stimulated tremendous effort to search for large-gap 2D TIs. The main strategy is to look for systems containing heavy elements, because the SOC effect, which is intimately connected with nontrivial band topology, typically increases with the atomic number. As the heaviest nonradioactive element, Bi has received much attention, and it is the common constituent of many well-known TIs discovered to date. Particularly, large-gap 2D TI has been found in the single Bi(111)-BL structure, which is a buckled honeycomb structure of Bi sheet, similar to the graphene. Its signature of TI with the well-localized topological edge states have been successfully verified by several experiments. The theoretical band gap of free standing Bi(111)-BL is ~0.12 eV (Fig. 1a), which is large enough for measurements and device applications[29]. Thus, the Bi(111)-BL system can serve as a paradigmatic model system for studying QSH effect.

In this work, we shall investigate the vdW heterostructure formed by Bi(111)-BL and the 2D FE material α-$In_2Se_3$. Fig. 1(c) and (d) show the structures of the 2D Bi(111)-BL/α-$In_2Se_3$ vdW heterostructure. The Bi(111)-BL and monolayer (ML) of α-$In_2Se_3$ are both hexagonal crystal systems with similar lattice constants. In the α-$In_2Se_3$ monolayer, one monolayer contains five layers of atoms arranged in the order of Se-In-Se-In-Se. The position of the middle layer of Se atoms determines the electric polarization direction in α-$In_2Se_3$. As shown in Fig. 1(a), three high symmetry stacking structures of Bi(111)-BL/α-$In_2Se_3$ are tested for both polarization directions, where the lower layer Bi atom is on-top of Se (①), In (②), and hollow (③) sites, respectively. The stability of these heterostructures can be inferred from the binding energy ($E_b$), defined as:

$$E_b = E(Bi) + E(In_2Se_3) - E(\text{heterostructure}), \quad (1)$$

where $E$ is the total energy. It is noted that the stacking sequence of Bi(111) BL on α-In$_2$Se$_3$ is not affected by the polarization. The most stable stacking structure is the same for both polarization direction with the lower Bi on-top the hollow site, namely, position ③ (see Tab. 1.). Meanwhile, the values of $E_b$ are quite different for different polarization directions: 0.52 and 0.68 eV for up and down polarizations, respectively. The interlayer distance between the two materials also changes from 2.82 to 2.57 Å (Fig. 1c and d) when the polarization is turned from up to down. Here, the up and down represent the polarization toward and away from Bi(111)-BL as shown in Fig. 1(c) and (d), respectively. The greater $E_b$ and the shorter interlayer distance indicates the interaction between Bi and In$_2$Se$_3$ become stronger when the polarization direction switches from up to down direction. This also greatly affects the electronic band structure of the heterostructure, which will be discussed below.

The free-standing Bi(111)-BL is inversion symmetric, and all bands are doubly degenerate (due to the combined spacetime inversion symmetry) with a global bandgap ~0.12 eV. For an inversion symmetric system, the $Z_2$ topological number can be determined from the parity eigenvalues for occupied bands at the time-reversal-invariant momentum (TRIM) points. The four TRIM points for Bi(111)-BL are the Γ point and the three $M$ points. The parity eigenvalues at these points are given in Fig. 2(a), and the obtained $Z_2$ number is clearly nontrivial. This shows that a free-standing Bi(111)-BL is a 2D TI, consistent with previous works.

After stacking on α-In$_2$Se$_3$ and forming a heterostructure, the inversion symmetry is broken and each twofold degenerated band will generally split except for the states at those high symmetric points, as shown in Fig. 2(b) and (c). Strong hybridization occurs between states from Bi(111)-BL and α-In$_2$Se$_3$ as their outer shell states are both *p*-orbitals. The hybridization mainly localizes at the Γ point of Brillion Zone, indicating this effect is delocalized and consistent with charge-transfer interaction[31] between them without significant covalent bonding. However, the states around Fermi level are quite different for two polarization directions of α-In$_2$Se$_3$, which is important for the topological property of the whole heterostructure.

As there is no inversion symmetry for the Bi(111)-BL/α-In$_2$Se$_3$ heterostructure, the topology cannot be directly obtained from the parity eigenvalues. In order to study the effects of α-In$_2$Se$_3$ on the topological property of Bi(111)-BL, we first plot the band structures (Fig. 3) of Bi(111)-BL, Bi(111)-BL/α-In$_2$Se$_3$-up and -down with the projection onto Bi-$s$, Bi-$p_z$ and Se-$p_z$ orbitals, where the size of the circles is proportional to the contribution of orbitals. [Other orbitals are less important for the low-energy states (Fig. S3 and Fig. S4).] Similar to the free-standing Bi, the possible band inversion should occur around the Γ point, so we may focus on the states at this point. In Bi, the four low energy states (with decreasing energy) around the Fermi level belong to the irreducible representations of $\Gamma_4^-$, $\Gamma_5^+ + \Gamma_6^+$, $\Gamma_4^-$ and $\Gamma_4^+$ for D$_{3d}$ point group, and the Fermi level lies between $\Gamma_5^+ + \Gamma_6^+$ and $\Gamma_4^-$ (see Fig. 3(a)). Now, by observing the orbital projection weights, we can trace the evolution of these states from Bi in the band structures of the combined system. The combined system of Bi(111)-BL/α-In$_2$Se$_3$ has a reduced symmetry of C$_{3v}$. Correspondingly, the representations $\Gamma_4^-$, $\Gamma_5^+ + \Gamma_6^+$, $\Gamma_4^-$ and $\Gamma_4^+$ evolves into $\Gamma_4$, $\Gamma_5 + \Gamma_6$, $\Gamma_4$ and $\Gamma_4$, respectively, and these Bi-dominated states are marked in Fig. 3(b) and 3(c). By comparing Fig. 3(a) and 3(b), one can observe that the band topology of Bi(111)-BL/α-In$_2$Se$_3$-up is the same as the free-standing Bi(111)-BL. Note that in the heterostructure, we can no longer distinguish the original $\Gamma_4^-$ and $\Gamma_4^+$ states, as the inversion is broken. But what is important for the topology here is the band ordering across the band gap, determined by the relative positions of these states. On the other hand, by comparing Fig. 3(a) and 3(c), one observes that the situation is different: for Bi(111)-BL/α-In$_2$Se$_3$-down, three states $\Gamma_4$, $\Gamma_5 + \Gamma_6$, and $\Gamma_4$ are above the Fermi energy, and only one $\Gamma_4$ is below the Fermi energy, in sharp contrast with the other two cases. The above analysis suggests that: Bi(111)-BL/α-In$_2$Se$_3$-up should have the same nontrivial topology as Bi(111)-BL, i.e., it remains a 2D TI; however, Bi(111)-BL/α-In$_2$Se$_3$-down may have a distinct topological character from Bi(111)-BL, i.e., it may be topologically trivial. Thus, the switching of the electric polarization of α-In$_2$Se$_3$ may drive a topological phase transition in the Bi(111)-BL.

To confirm the topological property, we explicitly compute the $Z_2$ number by using the Wilson loop method which traces the Wannier center evolution for the valence bands. (Note that for Bi(111)-BL/α-In$_2$Se$_3$-down, although its band structure has no global band gap, the local gap exists at every point in the Brillouin zone, so the $Z_2$ number is still well defined.[32]) As shown in Fig. S5, the obtained $Z_2$ number is 1 and 0 for the up and down polarization directions respectively. This confirms that Bi(111)-BL/α-In$_2$Se$_3$-up is still a QSH insulator, whereas Bi(111)-BL/α-In$_2$Se$_3$-down becomes a trivial system. α-In$_2$Se$_3$ is 2D ferroelectric material with two stable polarized states which can be switched from one to the other by the application of an external electric pulse, and these states retain robust even if the external field is removed[33-34]. Therefore, the nonvolatile electrical control of topological states can indeed be realized in 2D Bi(111)-BL/α-In$_2$Se$_3$ vdW heterostructures.

The evolution of Bi-$p$ orbitals around the Γ point can be summarized in a schematic diagram in Fig. 4(a)-(c). The bonding and antibonding Bi-$p$ orbitals split into $p_{x,y}$ and $p_z$ orbitals in this 2D honeycomb crystal field which hybridizes $p_x$ and $p_y$ orbitals in the 2D plane and leaves the $p_z$ orbitals out of plane. These states are labelled as $p^-_{x,y}$, $p^+_{x,y}$, $p^-_z$ and $p^+_z$ where the superscripts +(-) stands for even(odd) parity of the corresponding states. Because of the strong SOC effect of Bi, the two states with different parity, $p^+_{x,y}$ and $p^-_z$, are inverted and the odd state $p^-_z$ shift across the Fermi level, resulting in topological nontrivial character (Fig. 4a). When Bi(111)-BL is put on α-In$_2$Se$_3$, the hybridization at Γ point mainly occurs between Bi-$p_z$ and Se-$p_z$ orbitals. In Bi(111)-BL/α-In$_2$Se$_3$-up heterostructure, it occurs far away from Fermi level, so the ordering of Bi states is maintained. Hence, the up polarization does not change the topological property of Bi(111)-BL. When the polarization switches to the down direction, the charge transfer from Bi-$p_z$ and Se-$p_z$ orbitals makes $p^-_z$ state move up across the Fermi level, resulting in the change of the topological property of Bi(111)-BL. This charge transfer picture is consistent with the differential charge density between Bi(111)-BL and α-In$_2$Se$_3$. As shown in Fig. 4(d) and (e), yellow and blue colors represent electron gain and loss regions respectively. There is a little charge transfer between Bi(111)-BL and In$_2$Se$_3$ with up polarization while significant electron transfer occurs from Bi to the neighboring Se atom of α-In$_2$Se$_3$ with the down polarization, which is similar to the graphene/α-In$_2$Se$_3$ vdW heterostructure[35].

The distinct charge transfer behavior of the two polarization states can be understood by the polarization-dependent electrostatic potential difference between Bi(111)-BL and α-In$_2$Se$_3$ [36]. The in-plane average electrostatic potentials of Bi(111)-BL and α-In$_2$Se$_3$ are shown in Fig. 5(a) and 5(b). The Bi(111)-BL is a symmetric system with only one electrostatic potential while α-In$_2$Se$_3$ has an asymmetric structure along the out-of-plane direction thus its electrostatic potential differ significantly for the two sides. This difference converges at ~1.4 eV resulting in different band alignments when contacting Bi(111)-BL. In Fig. 5 (c), all energy levels are aligned by setting the vacuum level ($E_{VAC}$) to zero. For the up polarization, the valence band maximum (VBM) of Bi(111)-BL is lower than the conduction band minimum (CBM), and much higher than VBM of α-In$_2$Se$_3$ (left part of Fig. 5c), causing little charge transfer between them. It follows that the topology remains as that of freestanding Bi(111)-BL. On the other hand, the valence and conduction bands of α-In$_2$Se$_3$ move down in energy and CBM becomes much lower than VBM of Bi(111)-BL when the polarization switches to the down direction (right part of Fig. 5c). Accordingly, the electrons transfer from Bi(111)-BL to α-In$_2$Se$_3$ makes one $\Gamma_4^-$ from Bi move across the Fermi level (Fig. 3c). This evolution changes the topology. Thus, the charge transfer controlled by the reversible polarization plays an important role in controlling the topological character of Bi(111)-BL/α-In$_2$Se$_3$.

The hallmark of a QSH insulator is the existence of topologically protected conducting states at edges in the bulk band gap[37-38]. Here, we note that the global band gap in Bi(111)-BL/α-In$_2$Se$_3$-down vanishes. To better visualize the edge states, we can apply a moderate tensile strain (~2%) to produce a positive band gap. As shown in Fig. S6, the band structures are very similar to the unstrained ones except for a finite global gap. We also checked that their topological properties are the same. To study the edge states, in Fig. 6, we plot the spectra for zigzag nano ribbons (ZNR) of Bi(111)-BL/α-In2Se3-up and -down. The width of ribbons is about 6.5 nm. One observes that the Bi(111)-BL/α-In$_2$Se$_3$-up ribbon has two pairs of Dirac like bands in the bulk gap. By checking their wave function distributions, we confirm that the two pairs are located at the two edges of the ribbon. The energy splitting between them is because the two edges of ribbon are not equivalent due to the substrate. Meanwhile, for the down polarization direction, no edge states appear in the bulk band gap. These results perfectly agree with our previous analysis, and it further confirms the topological switching by the FE polarization of α-In$_2$Se$_3$.

We have further investigated the heterostructures with more layers of α-In$_2$Se$_3$ and Bi(111)-BLs. It is found that the features of Bi(111)-BL on two layers of α-In$_2$Se$_3$ is very similar to the above results of Bi(111)-BL/α-In$_2$Se$_3$ (Fig. S8). The Bi(111)-BL on two layers of α-In$_2$Se$_3$ with up polarization still remains a 2D TI. And this heterostructure become a trivial metal with down polarization. Since the polarization value of α-In$_2$Se$_3$ is already saturated at two layers, it is expected that the band topology of Bi(111)-BL does not change further with more α-In$_2$Se$_3$ layers.

For practical applications, a finite band gap is desired for both states. In this regard, Bi(111)-BL/α-In$_2$Se$_3$ is not an ideal system, because the Bi(111)-BL/α-In$_2$Se$_3$-down state does not have a global band gap. As we have shown, a possible way to tackle this problem is to apply a moderate lattice strain, which can open a global gap in Bi(111)-BL/α-In$_2$Se$_3$-down. It should also be stressed that the purpose (and the significance) of this work is to demonstrate the novel concept of topological switching using 2D FE polarizations. The Bi(111)-BL/α-In$_2$Se$_3$ serves its purpose to demonstrate the idea. In view of the rapidly expanding family of 2D TIs and FE materials, more suitable vdW heterostructures realizing our proposed concept can be expected in the near future.

## 4 Conclusion

In conclusion, we propose a new approach to achieve topological switching in 2D TI via FE polarizations. We demonstrate this idea via a concrete vdW heterostructures consisting of Bi(111)-BL and 2D FE α-In$_2$Se$_3$ monolayer. The topological nontrivial/trivial states can be switched by the reversal of the electric polarization of α-In$_2$Se$_3$. The topological switching has been verified from various aspects. We show that the charge transfer between Bi(111)-BL and α-In$_2$Se$_3$ plays an important role in the switching mechanism. Compared with previous proposals, the new approach has the unique advantage of non-volatility. This work thus opens a new route to realize controlled topological states, which has promising potential applications in topological electronics devices, especially for the non-volatile devices.

**Acknowledgments**

The authors thank Lan Chen for discussions. This work was supported by the National Natural Science Foundation of China (Grant No. 11974307, 11790313, 11774305) and the National Key Projects for Research & Development of China (Grant Nos. 2016YFA0300402 and 2019YFA0308602), Zhejiang Provincial Natural Science Foundation (D19A040001), the Fundamental Research Funds for the Central Universities and the Singapore MOE AcRF Tier 2 (MOE2019-T2-1-001)

**Conflict of interest**

*The authors declare no conflict of interest.*

# Tables

**Tab. 1.** The binding energy of Bi(111)-BL on α-In$_2$Se$_3$ with different polarization direction.

| $E_b$ (eV) | top Se | top In | hollow |
|---|---|---|---|
| up | 0.48 | 0.65 | 0.68 |
| down | 0.39 | 0.5 | 0.52 |

**Figures:**

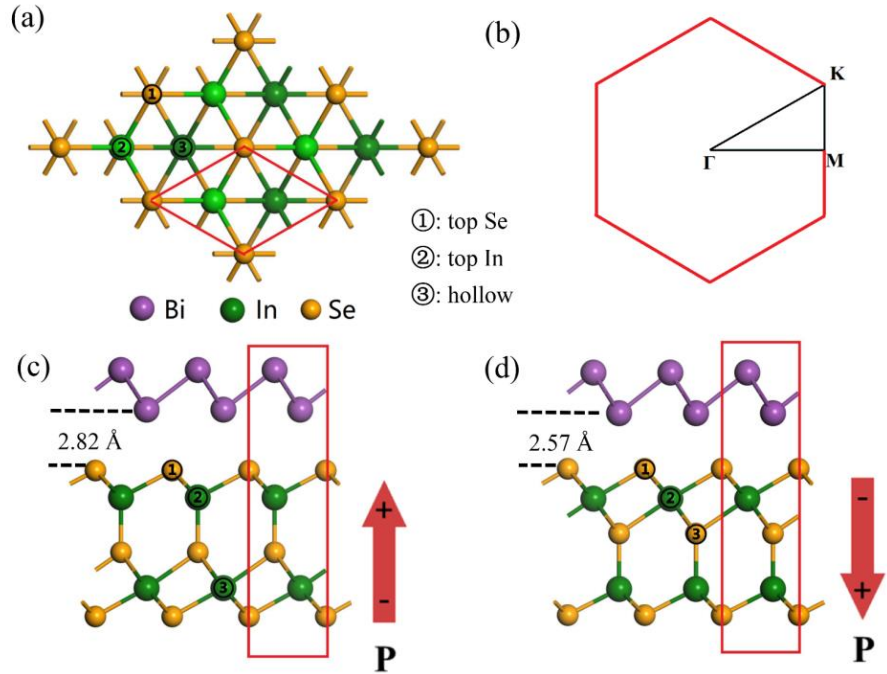

**FIG. 1.** (a) α-In$_2$Se$_3$ at the top view. In order to distinguish the top and bottom In, we increase the brightness of top In. (b) The Brillium Zone of 2D hexagonal system. (c) Bi(111)-BL/α-In$_2$Se$_3$-up and (d) Bi(111)/α-In$_2$Se$_3$-down at the side view. Sites ① - ③ represent different adsorption top sites for the bottom Bi layer. Red lines indicate the unit cell for calculations. Atoms with purple, orange and green colors stand for bismuth, selenium and indium, respectively. Red arrows indicate the electric polarization direction of In$_2$Se$_3$.

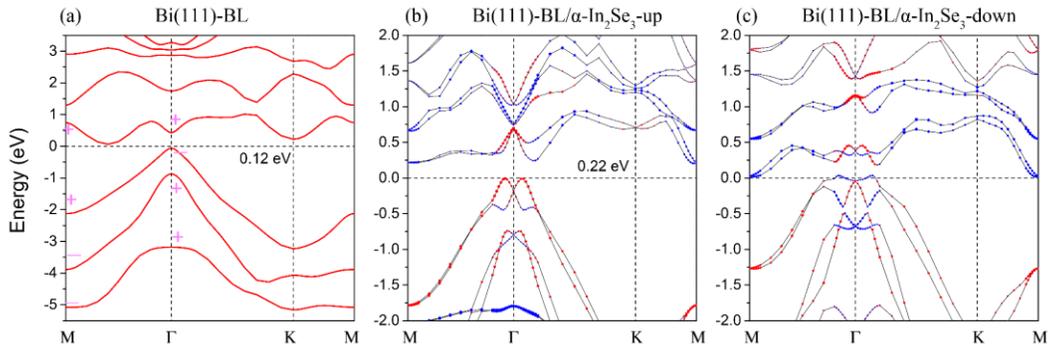

**FIG. 2.** Band structures of Bi(111)-BL (a), Bi(111)-BL/α-In₂Se₃-up (b) and -down (c). Red and blue lines identify the bands of Bi and In₂Se₃ respectively. The signs in (a) at Γ and M points are parity eigenvalues.

**FIG. 3.** The projected orbital components of Bi-$s$, Bi-$p_z$, and Se-$p_z$: (a) (d) Bi(111)-BL (g) In$_2$Se$_3$-ML (b) (e) (h) Bi(111)-BL/α-In$_2$Se$_3$-up. (c) (f) (i) Bi(111)-BL/α-In$_2$Se$_3$-down. The irreducible representations of Bi at Γ points were labeled in (a) - (c).

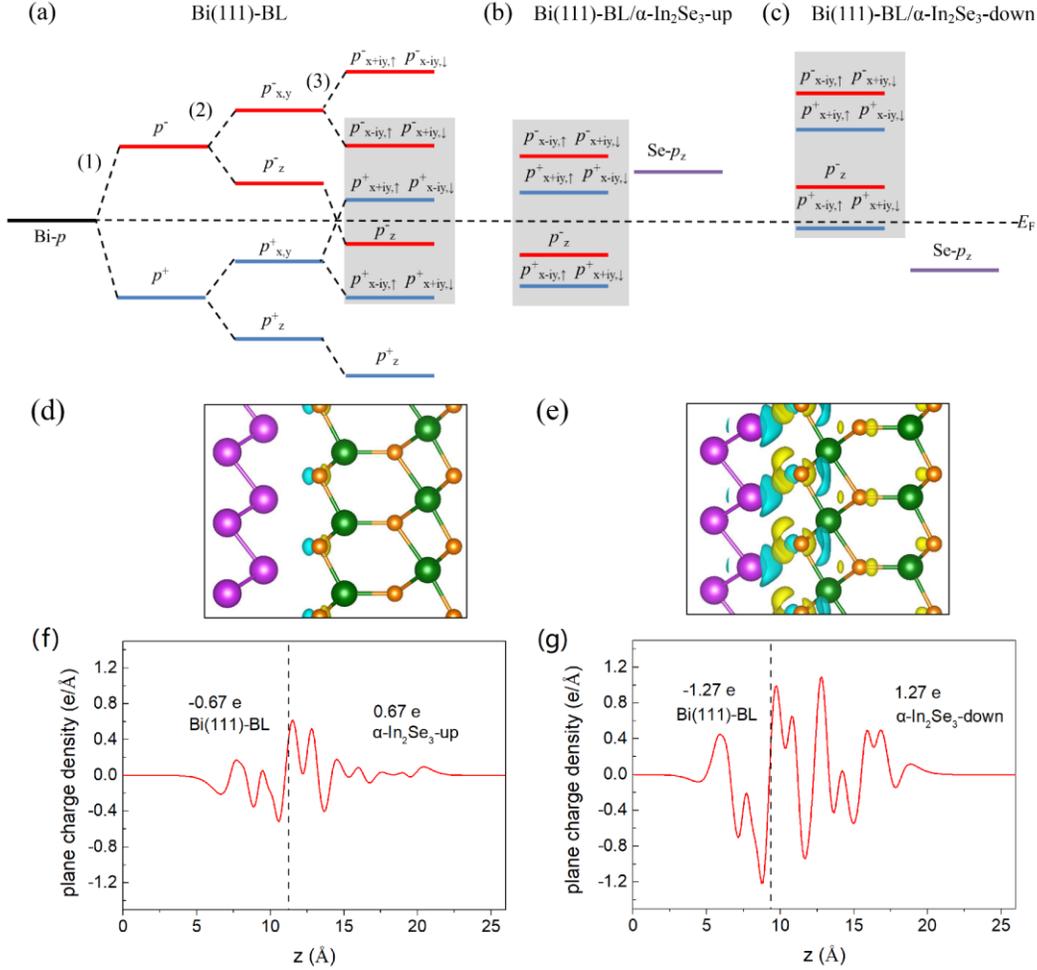

**FIG. 4.** (a) - (c) Schematic diagram of the evolution from the orbitals at Γ point in the three systems. In heterostructures we concentrate on the four states of Bi mentioned above. (1), (2), (3) in (a) are the effects of chemical bonding, crystal-field splitting and spin-orbit coupling, respectively. (d) (e) The differential charge density of Bi(111)-BL/α-In$_2$Se$_3$-up and -down. (f) (g) Corresponding plane charge density along z-axis. The numbers in pictures are estimated values of charge transfer.

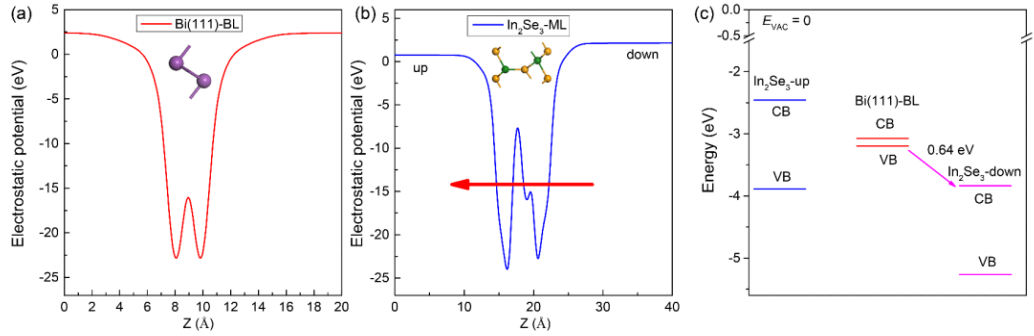

**FIG. 5.** (a) and (b) Electrostatic potential of Bi(111)-BL and In$_2$Se$_3$-ML respectively. The red arrow in (b) indicate the polarization direction of In$_2$Se$_3$. (c) Band alignments between Bi(111)-BL and In$_2$Se$_3$-ML with different polarization directions. All the energy levels are shifted relative to the vacuum level.

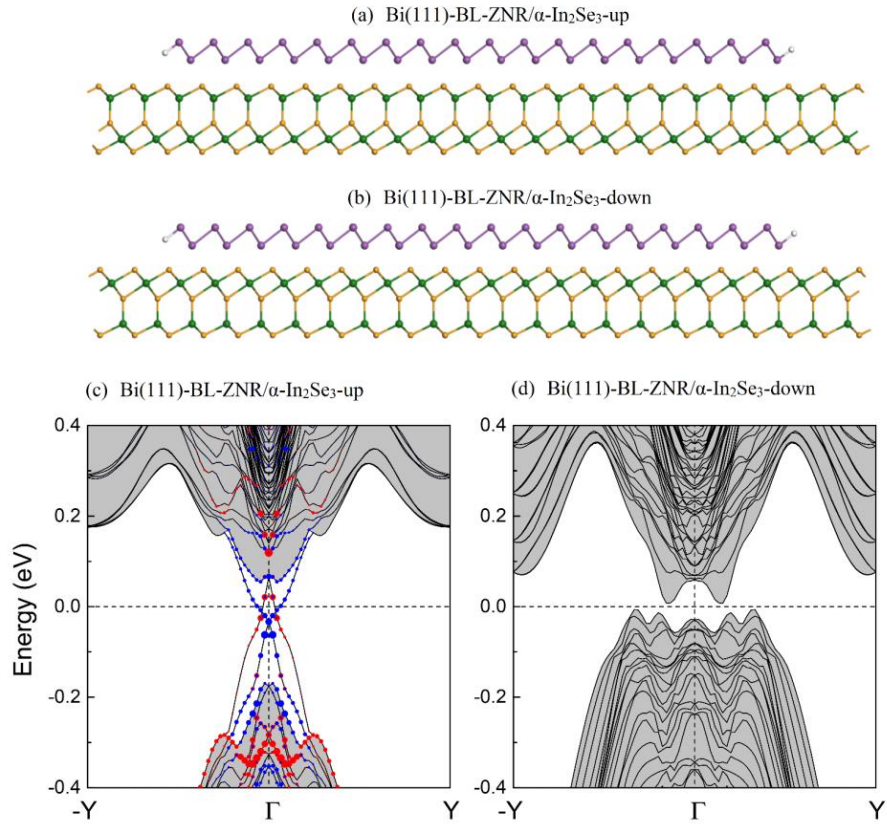

**FIG. 6.** (a) (b) Side view of Bi(111)-BL-ZNR/α-In$_2$Se$_3$-up and -down. (c) (d) Band structures of Bi(111)-BL-ZNR/α-In$_2$Se$_3$-up and -down with -2% biax strain. The red and blue dots in (c) are the projections for left edge and right edge respectively. The bulk states are indicated by shadow.

# Electronic Supplementary Material

# Nonvolatile ferroelectric control of topological states in 2D heterostructures

Hua Bai[1], Xinwei Wang[2], Weikang Wu[3], Pimo He[1], Zhu'an Xu[1], Shengyuan A. Yang[3] and Yunhao Lu[1] (✉)

[1] Zhejiang Province Key Laboratory of Quantum Technology and Device, Department of Physics, Zhejiang University, Hangzhou 310027, China

[2] State Key Laboratory of Silicon Materials and School of Materials Science and Engineering, Zhejiang University, Hangzhou 310027, P. R. China

[3] Research Laboratory for Quantum Materials, Singapore University of Technology and Design, Singapore 487372, Singapore

Address correspondence to Yunhao Lu: luyh@zju.edu.cn

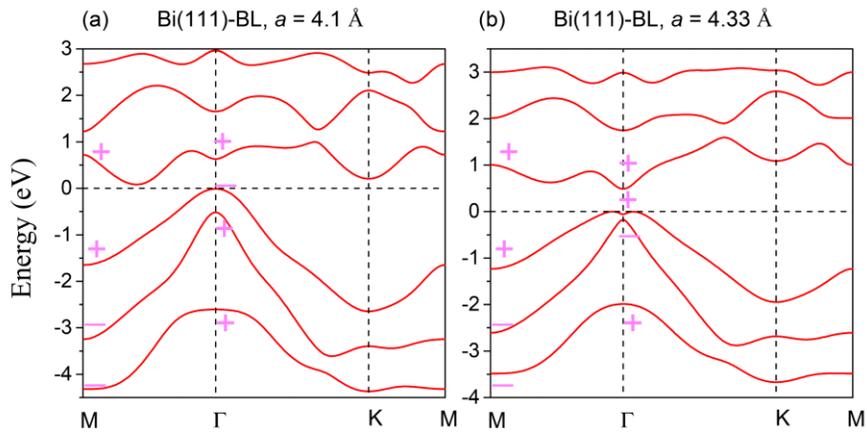

**FIG. S1.** PBE band structures of Bi(111)-BL with a = 4.1 Å (a) and 4.33 Å (b).

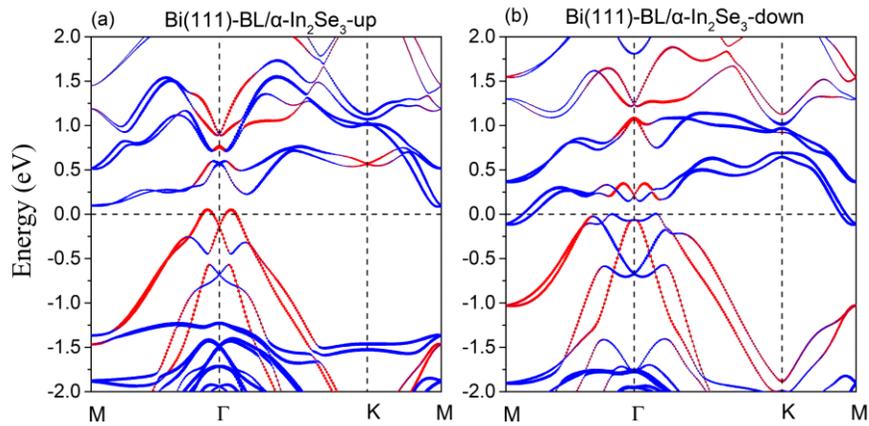

**S2.** PBE band structures of Bi(111)-BL/α-In$_2$Se$_3$-up (a) and -down (b). Red and blue lines identify the bands of Bi and In$_2$Se$_3$ respectively.

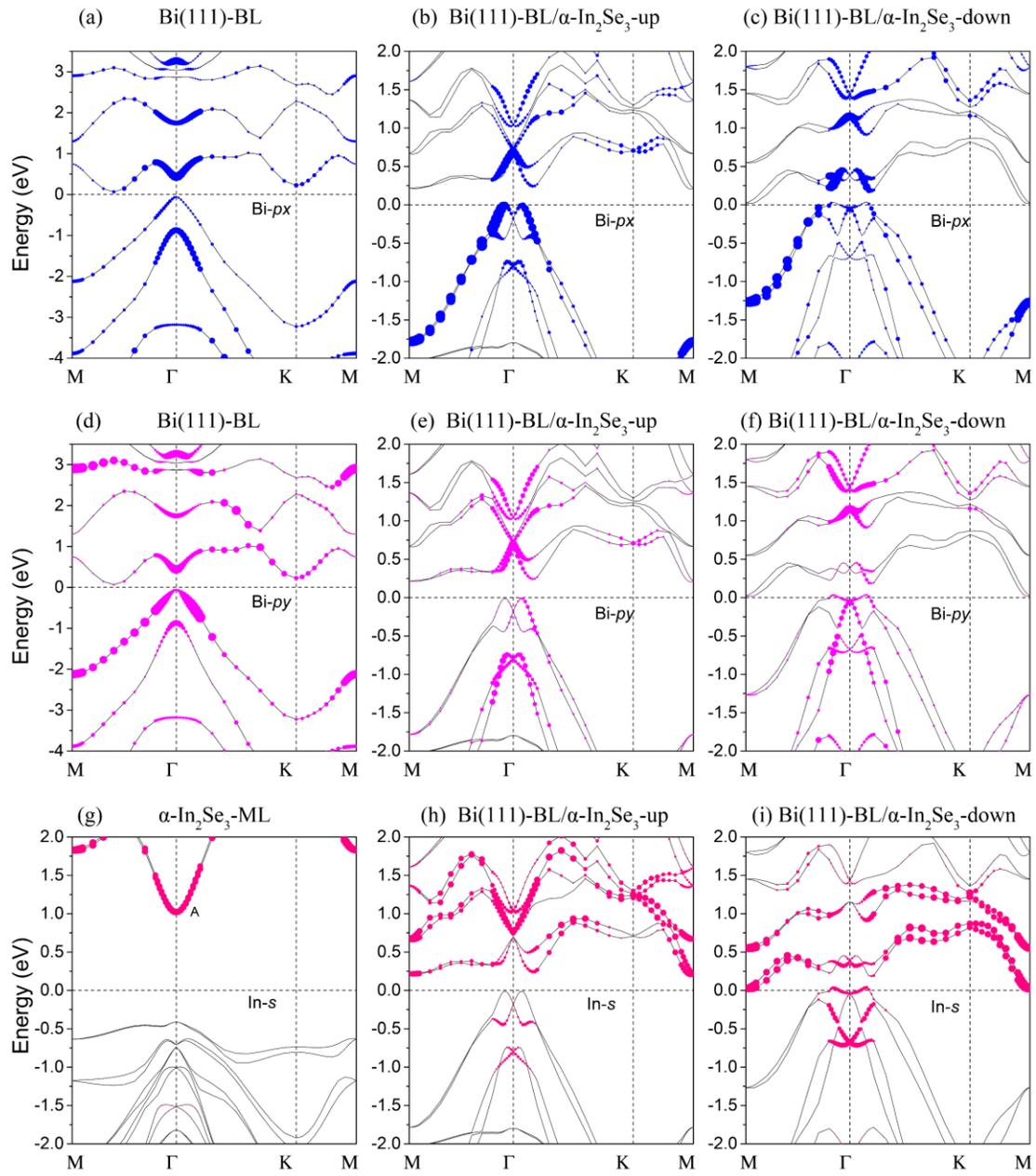

**FIG. S3.** The projected orbital components of Bi- $p_x$, Bi-$p_y$, and In-$s$: (a) (d) Bi(111)-BL (g) In$_2$Se$_3$-ML (b) (e) (h) Bi(111)-BL/α-In$_2$Se$_3$-up. (c) (f) (i) Bi(111)-BL/α-In$_2$Se$_3$-down.

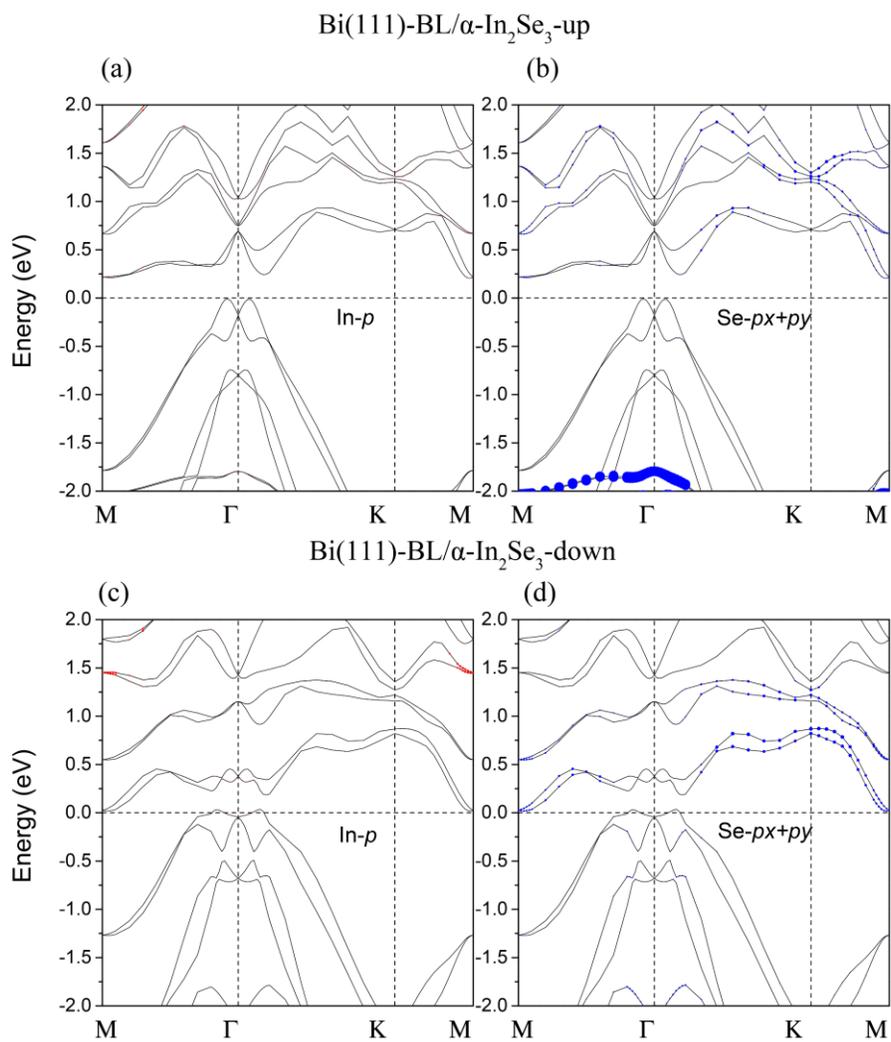

**FIG. S4.** The projected orbital components of and In-$p$ and Se-$p_x$+$p_y$ for the two system.

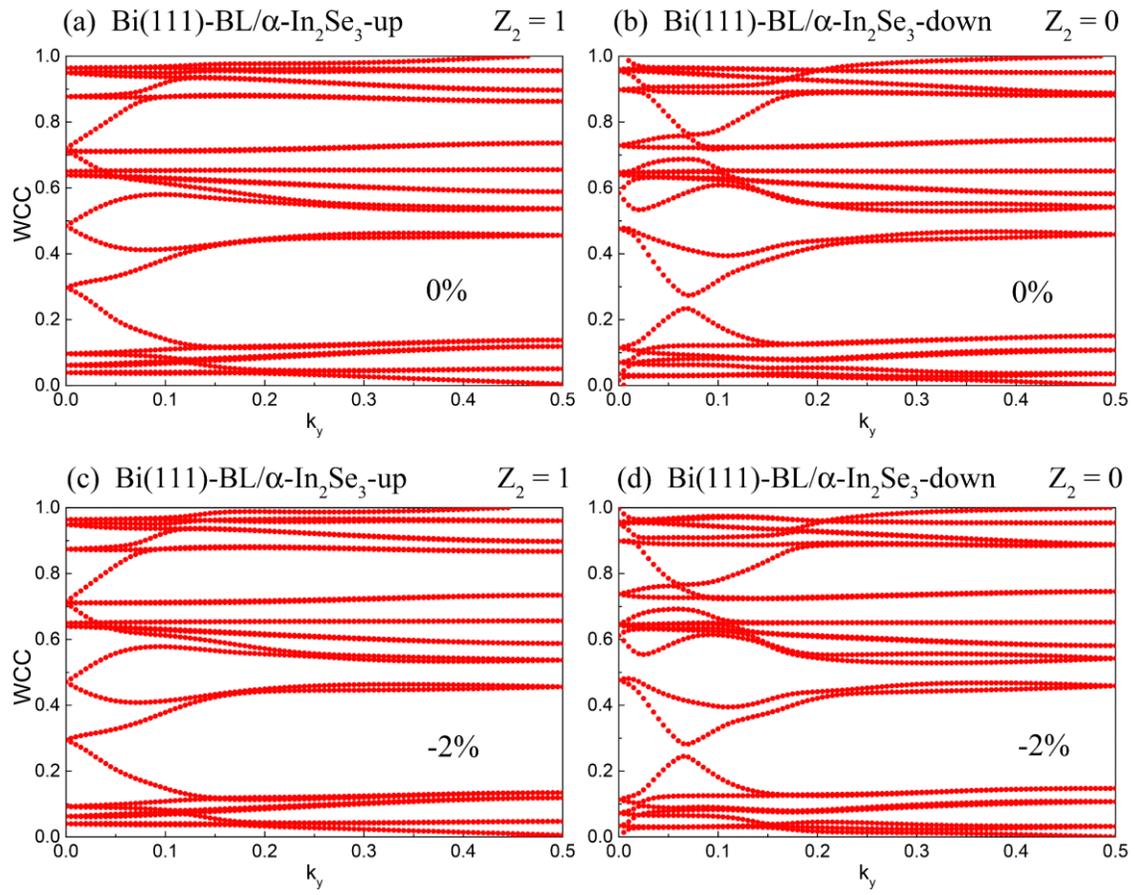

**FIG. S5.** Wannier charge center (WCC) evolution of Bi(111)-BL/α-In$_2$Se$_3$-up and –down with 0% (a, b) and -2% strain (c, d).

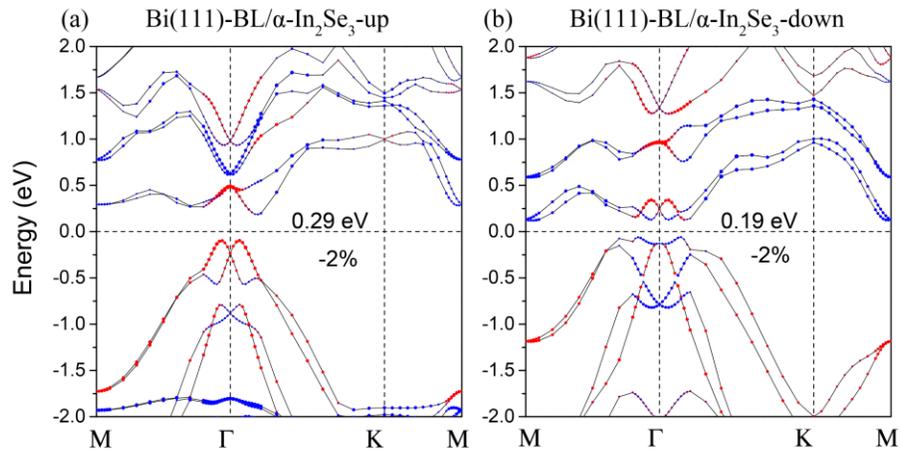

**FIG. S6.** Band structures of Bi(111)-BL/α-In$_2$Se$_3$-up (a) and -down (b) with -2% biax strain.

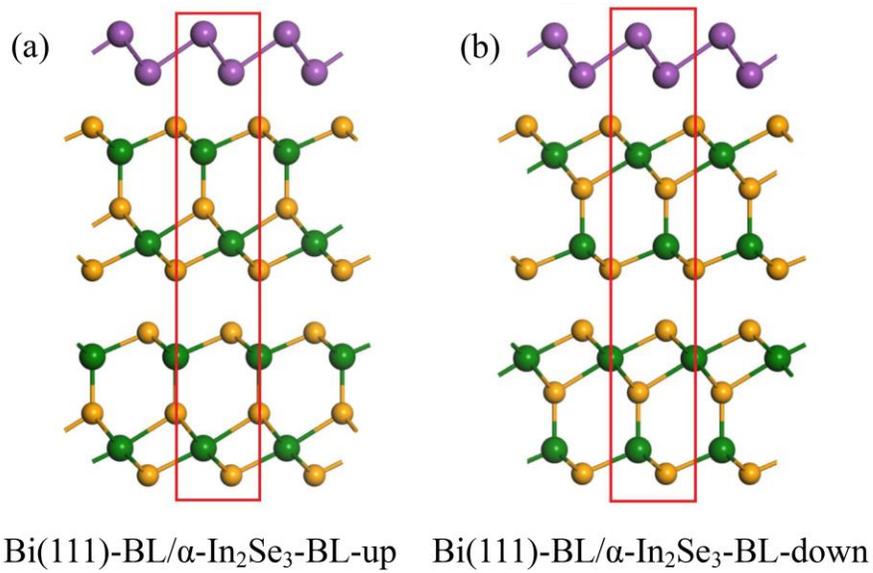

Bi(111)-BL/α-In$_2$Se$_3$-BL-up    Bi(111)-BL/α-In$_2$Se$_3$-BL-down

**FIG. S7.** The side view of Bi(111)-BL/α-In$_2$Se$_3$-BL-up (a), -down (b).

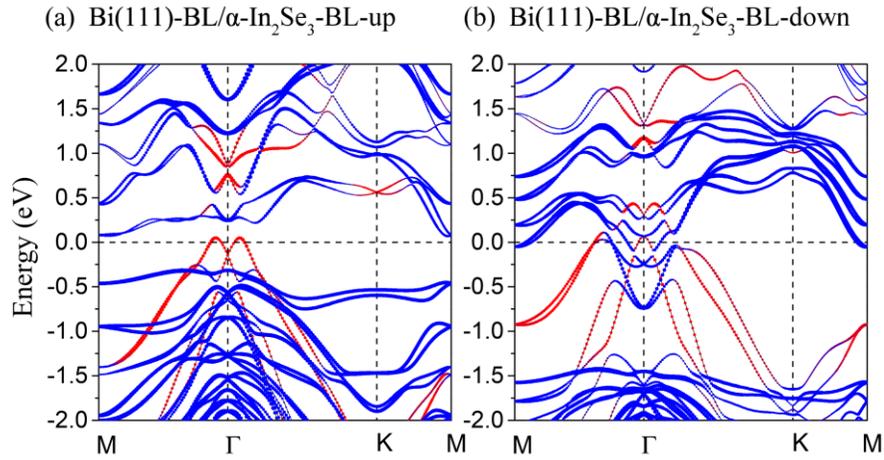

**FIG. S8.** PBE band structures of Bi(111)-BL/α-In2Se3-BL-up (a), -down (b) and Bi(111)-2BLs/α-In2Se3-up (c), -down (d). Red and blue lines identify the bands of Bi and In$_2$Se$_3$ respectively.

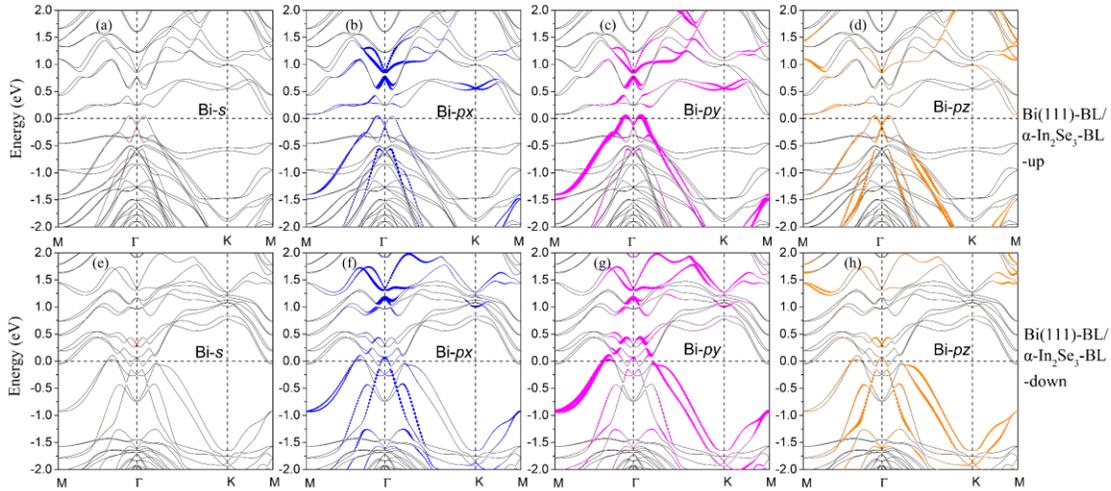

**FIG. S9.** The projected orbital components of Bi in three systems: (a) - (d) Bi(111)-BL/α-In2Se3-BL -up. (e) - (h) Bi(111)-BL/α-In2Se3-BL-down for PBE bands.

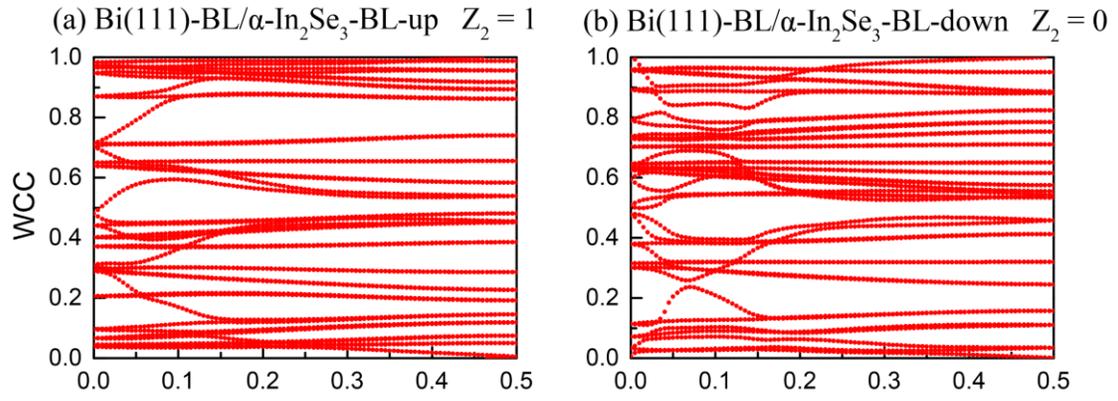

**FIG. S10.** WCC evolution of Bi(111)-BL/α-In$_2$Se$_3$-BL-up and -down.